# A Novel Option for Waste Tire Rubber Reutilization: Refrigerant in Solid-State Cooling Devices

Nicolau Molina Bom,[a,b] Érik Oda Usuda,[c,d] Mariana da Silva Gigliotti,[c] Denílson José Marcolino de Aguiar,[c,e] William Imamura,[c,f] Lucas Soares Paixão,[c] and Alexandre Magnus Gomes Carvalho[*c,f,g]

[a]*Catalan Institute of Nanoscience and Nanotechnology (ICN2), 08860, Belaterra, Barcelona, Spain*
[b]*The Institute of Photonic Sciences (ICFO), 08193, Castelldefels, Barcelona, Spain*
[c]*Laboratório Nacional de Luz Síncrotron (LNLS), Centro Nacional de Pesquisa em Energia e Materiais (CNPEM),13083-100, Campinas, SP, Brazil*
[d]*Universidade Federal de São Paulo (UNIFESP), 00972-270, Diadema, SP, Brazil.*
[e]*Universidade tecnológica Federal do Paraná (UTFPR), 80230-901, Ponta Grossa, PR, Brazil*
[f]*Faculdade de Engenharia Mecânica, Universidade Estadual de Campinas (UNICAMP), 13083-860, Campinas, SP, Brazil.*
[g]*Departamento de Engenharia Mecânica, UEM, 870-900, Maringá, PR, Brazil*
\* E-mail: alexandre.carvalho@lnls.br

**ABSTRACT**

Management of discarded tires is a compelling environmental issue worldwide. Although several approaches have been developed to recycle waste tire rubbers, their application in solid-state cooling is still unexplored. Considering the high barocaloric potential verified for elastomers, the use of waste tire rubber (WTR) as refrigerant in solid-state cooling devices is very promising. Here, we investigated the barocaloric effects in WTR and polymer blends made of vulcanized natural rubber (VNR) and WTR, in order to evaluate its feasibility for solid-state cooling technologies. The adiabatic temperature change and the isothermal entropy change reach giant values, as well as the performance parameters, being comparable or even better than most barocaloric materials in literature. Moreover, pure WTR and WTR-based samples also present a faster thermal exchange than VNR, consisting in an additional advantage of using these discarded materials. Thus, the present findings evidence the encouraging perspectives of employing waste rubbers in solid-state cooling based on barocaloric effect, contributing in both the recycling of polymers and the sustainable energy technology field.

# INTRODUCTION

Polymers are widespread in our everyday life, with applications in practically all branches of science and technology. For this reason, the term "Plastic Age" is sometimes used to refer to the recent period. This increasing amount of plastics in industrial products creates a serious ecological issue: without proper management of wastes, the long durability of these materials results in their accumulation in nature. Besides, the microplastic particles released by the polymer decomposition are also environmentally harmful.[1] Most of this polymeric material worldwide is used in automobile tires: estimates indicate that 1.5 billion of them are manufactured each year,[2,3] 800 million are discarded and this number is expected to grow 2% annually.[4] A tire is a complex engineering product, mainly composed by elastomers, but also containing textiles and metals.[5] The production of crosslinked elastomers involves the vulcanization process, which prevents this class of polymers to be completely recycled or reprocessed, on the contrary of what happens with thermoplastics.[6] Considering the growing environmental awareness and the possibility of reusing valuable raw materials contained in tires, end-of-life strategies for management of used tires are highly demanding.

The classified methods for waste tire management of rubbery materials are: landfill, energy recovery, recycling, re-use and prevention;[7] following the European Union (EU) hierarchy,[8] from the least to the most desirable. Although land filling is the easiest way to discard used tires, the hazardous character of rubber waste makes this practice unacceptable in view of the current ecological requirements, being prohibited by EU since 1999.[9] Energy recovery is the second method for rubber waste management, since tires present a caloric value equivalent or higher than coal.[10] Re-use of scrap tires by retreading is the most economically viable method for waste tire utilization, but its market is still scarce.[10,11] Recycling is the most used approach for old tire management, and should be preferred to recovery, since only 30-38% of the energy contained in a new tire can be recovered.[12,13] To date, several recycling methods are being employed,[6] most of them involving the grinding of rubber waste in a powder.[14,15] Grinding rubber waste for recycling purposes was first proposed by Goodyear in 1853.[16]

Among waste tire rubber (WTR) recycling approaches, blending the rubber waste particles with polymers has been shown to be very effective, allowing to reduce the costs of the derived products.[4,17] WTR polymer blends were already attempted for the three families of polymers: elastomers, thermosets and thermoplastics. In all cases, it is possible to obtain a final product with good mechanical characteristics, which can be further improved by reducing the particle size of rubber granulates[18,19] or enhancing the compatibilization between the blend matrix and WTR powder.[20–23] Despite this promising potential for recycling verified in WTR-based polymer blends, their properties in view of solid-state cooling are still unexplored.

The current refrigeration technology is based on vapor-compression cycles employing hazardous gases as refrigerants, which may present global warming potentials up to 2000 times that of $CO_2$.[24] Besides, the energetic efficiency of these machines is quite low, reaching only 30% of the Carnot efficiency.[25] Solid-state cooling based on $i$-caloric materials is considered the most promising approach to replace the conventional cooling systems, by using eco-friendly materials with higher energy efficiencies.[25–29] *i*-caloric effects are characterized by temperature and entropy changes induced by the application of external fields on a material, where *i* stands for intensive variables. Very recently, giant barocaloric effects, which are driven by the application of isostatic pressure, were observed in distinct elastomers: vulcanized natural rubber (VNR);[30,31] PDMS;[32] and acetoxy silicone rubber.[33] In all these cases, barocaloric effect values are higher than those observed in other classes of materials, such as intermetallics. Besides revealing the high barocaloric potential of the elastomers, these results also open an encouraging perspective for using waste tire rubber in such applications, since most of the typical tire composition is represented by elastomeric materials.[5] Moreover, the possibility of combining a novel recycling method with an energy saving technology – represented by solid-state cooling – make this approach particularly appealing in view of sustainability.

Taking into account the discussion above, in the present study, we systematically investigated the barocaloric effects in WTR and polymer blends composed by WTR and VNR. The experiments were performed in a customized experimental apparatus, previously developed by our group.[34] The results show the barocaloric effects (adiabatic temperature change and isothermal entropy change) and the performance parameters in WTR-based samples are close to those values obtained from pure VNR. Moreover, the thermal exchange is significantly faster in WTR than in VNR, constituting an additional gain in using waste rubbers.

**MATERIALS AND METHODS**

Samples

We prepared the VNR samples using commercial pre-vulcanized latex (purchased from Siquiplas). The 8-mm-diameter VNR cylinders were cast into a plaster mold, with a latex feed to prevent the formation of cavities during the drying process. The density of the samples is 902(7) kg m$^{-3}$, checked by a pycnometer.

WTR powder used in the blends was supplied by UTEP Company. The diameters of the granules are distributed from ~10 μm to ~3 mm, with the highest concentration within the 0-100 μm range (~56%), as displayed in Table A1 (Appendix).

Polymer blend samples were prepared by mixing ~33% of WTR powder to 67% of latex into a cylindrical steel mold. The material was pressed by a piston to remove part of the latex. By the height of the piston inside the mold, it is possible to roughly control how much of the latex is removed, neglecting the WTR powder going out with the liquid phase. In the final step, the blend is allowed to dry naturally for 3 days inside the mold and 7 days in air. The percentage of WTR in these 8-mm-diameter cylinders is indirectly determined by calculating the difference of the weights before and after the whole process.

WTR powder was also sintered into an 8-mm-diameter cylinder under pressure of 40 MPa for 30 min at high temperature of 453 K. This sample was only used for DSC measurement.

Fourier-transform Infrared Spectroscopy

Fourier transform infrared spectroscopy (FTIR) characterizations were performed from 500 to 4000 cm$^{-1}$ (fixed step of 2 cm$^{-1}$) using a FTIR spectrometer from PerkinElmer® (model Spectrum Two).

Fig. A1 (Appendix) displays the FTIR spectra for VNR, pure WTR and WTR 88 wt% blend. The peaks between 3100 and 2800 cm$^{-1}$, observed in all spectra, are attributed to C-H stretching vibration and are typical for VNR,[35] but are also verified in styrene-butadiene rubber compounds.[36] The peak at 1537 cm$^{-1}$, only appearing in the WTR-based samples, corresponds to the stretching vibration of double bonds in CH=CH groups, which is a clear indication of partial devulcanization during WTR production.[37] The presence of the peak at 1710 cm$^{-1}$, assigned to the absorption of carbonyl groups (C=O), results from the milling process.[37]

X-ray diffraction

X-ray diffraction (XRD) data was obtained at XRD1 beamline[38,39] at the Brazilian Synchrotron Light Laboratory (LNLS), using X-rays with energy of 12 keV.

The XRD profile in Fig. A2 (Appendix), corresponding to the pure WTR powder, displays the typical amorphous components, corresponding to elastomers usually contained in tires (natural and synthetic rubber, carbon black), but also exhibiting various crystalline peaks. It is worth mentioning WTR powder is formed by tires from different producers, which may vary considerably in compositional materials. Though, the presence of zinc compounds is expected, since zinc oxide (ZnO) is commonly used as an activator for the vulcanization reaction.[40] We also identify the presence of common rubber fillers such as calcium carbonate ($CaCO_3$) and silicon dioxide ($SiO_2$)002E.

Experiments under pressure

We performed the compression-decompression experiments in a piston-cylinder carbon-steel chamber surrounded by a copper coil, which enables the use of cooling/heating fluids (such as water and liquid nitrogen). A thermostatic bath (TE 184, Tecnal) was employed to reach temperatures above 280 K, by pumping water into the copper coil. Below 280 K, we used nitrogen to cool down the chamber. In both cases, two tubular heating elements (NP 38899, HG Resistências) were used to provide thermal stability to the system. A manual 15,000-kgf hydraulic press (P15500, Bonevau) was used to apply the uniaxial load. A load cell (3101C, ALFA Instrumentos) measured the contact force. A Cryogenic Temperature Controller (Model 335, Lake Shore Cryotronic) carried out the temperature acquisition and control. More details about the apparatus can be found in Bom et al.[34]

Direct measurements of barocaloric adiabatic temperature changes were performed in a four-step procedure: i) the sample is compressed quasi-adiabatically, causing a fast increase in temperature; ii) temperature cools down to the initial temperature, load is kept constant; iii) adiabatic decompression of the sample, causing an abrupt decrease in temperature; iv) sample temperature returns to the initial value. These cycles were performed only after temperature stabilization.

Specific heat from Differential Scanning Calorimetry

We determined the specific heat ($c_p$) as a function of temperature for the VNR, sintered WTR and the polymer blend samples (57, 73 and 88 wt%) using differential scanning calorimetry (DSC) data. The DSC measurements were carried out under ambient pressure, with heating rate of 10 K min$^{-1}$ from ~180 K to ~350 K.

Calculation of time constant of temperature vs. time curves

We used the following relationship, describing the variation of the temperature as a function of time, to fit the decompression behavior of our experimental curves:[41]

$$T(t) = T_0 + (T_1 - T_0)e^{-\frac{(t-t_1)}{\tau}} \quad (1)$$

where τ corresponds to the time constant; $T_0$ and $T_1$ are the initial and final temperatures, respectively; t is the time parameter and $t_1$ is the final time (when $T = T_1$). This model was fitted at the decompression region of temperature vs. time curves within the 333-233 K range (examples in Fig. A3 in Appendix). $T_0$ was defined as the inflection point of the exponential increase of temperature ($T_0 = T(\frac{d^2T}{dt^2} = 0)$), and $T_1$ was taken at the beginning of the temperature plateau. The data for 223 K were not included due phase transitions occurring at this temperature, precluding the fitting. The values obtained from this procedure are listed in Table A2 (Appendix).

**RESULTS AND DISCUSSION**

Barocaloric effects

Fig. 1 shows adiabatic temperature change ($\Delta T_S$) values obtained from the barocaloric experiments, measured in decompression, as a function of the distinct percentages of WTR, including pure VNR (0 wt%). These experiments were performed at the initial temperatures of 333, 293 and 243 K (Figs. 1a, b and c, respectively), for pressure changes ($\Delta p$) of 173(3), 260(8) and 390(12) MPa. One can verify a similar pattern in all experimental conditions, where $\Delta T_S$ consistently decreases as WTR content is increased. The large values of temperature change are mainly due to the natural rubber content. As the WTR content increases, the natural rubber content decreases giving place to others non-rubbery elements also present in the WTR, like oxides and other fillers (e.g. carbon black, fibers). Thus, it is expected that these other non-rubbery elements contribute insignificantly to the barocaloric effect. Therefore, the barocaloric effect tends to reduce when WTR content is increased when compared with the VNR. Nevertheless, the rate of decrease is quite low, even for samples with pure WTR powder (100 wt%). The maximum ΔT$_S$ reduction in comparison with pure VNR is 25%, obtained at 243 K for Δ$p$ = 260 MPa; the average ΔT$_S$ loss for the entire set of samples is ~21%. Although this reduction is significant, the ΔT$_S$ values measured in WRT-based samples are still within the range of giant barocaloric changes, suggesting the use of WTR powder as refrigerant in barocaloric devices is technically viable.

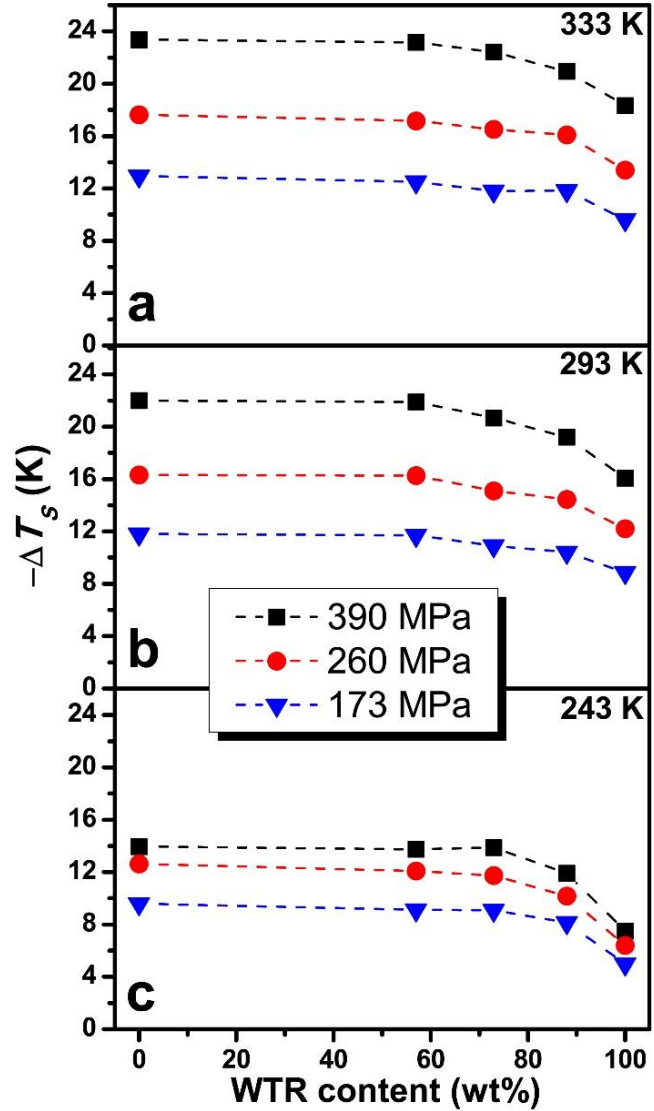

**Figure 1**. Direct measurements of ΔT$_S$ in decompression as a function of WTR weight percentages within the samples. The experiments were performed at the initial temperatures of 333 K (a), 293 K (b) and 243 K (c), for pressure changes of 173(3), 260(8) and 390(12) MPa. We estimate an error of 3% for pressure.

Furthermore, we investigated the $\Delta T_S$ of WTR-based samples, on decompression, within the 223-333 K temperature range ($\Delta p$ = 173, 260 and 390 MPa). Figs. 2a and 2b show $\Delta T_S$ as a function of initial temperature, corresponding to the WTR 88 and 100 wt% samples, respectively. Qualitatively speaking, both samples present the same trend: for lower temperatures, $\Delta T_S$ tends to decrease; above a temperature threshold, the curve changes its slope and $\Delta T_S$ varies in a lower rate. Moreover, analyzing $\Delta T_S$ x T curves, it is possible to notice this threshold shifts to higher temperatures as applied pressure increases. This behavior is consistent with the glass transition occurring in elastomers, already reported for VNR.[31] Below the glass transition temperature ($T_g$), the movements of polymer chains are largely limited, reducing the number of accessible states of the system, decreasing the barocaloric effect as a consequence. Considering WTR is mainly composed by elastomeric material, this mechanism can also explain the results in Fig. 2. Finally, $\Delta T_S$ values for the 88 wt% sample are consistently higher in comparison with the 100 wt% sample, reaching the maximum $\Delta T_S$ of ~21 K (for pressure change of 390 MPa at 323 K).

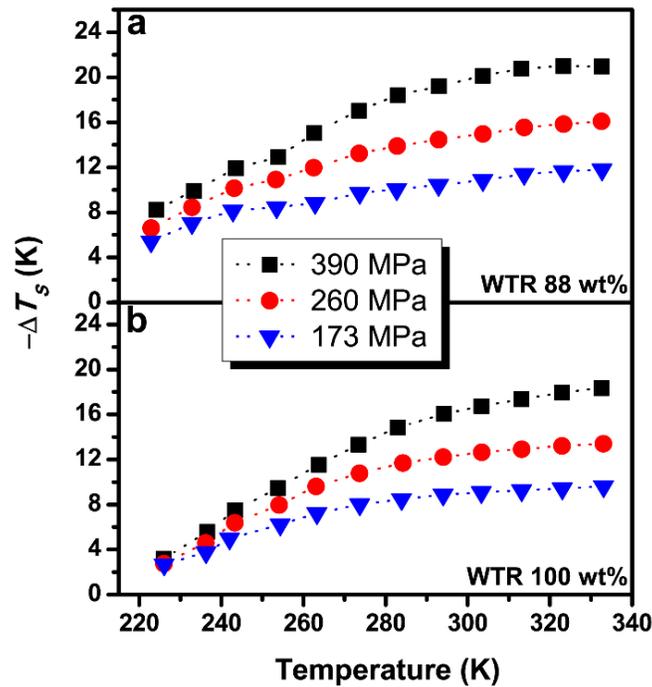

**Figure 2.** Direct measurements of $\Delta T_S$ in decompression as function of initial temperature, for samples with 88 wt% (a) and 100 wt% (b) of WTR. The experiments were performed within the 223-333 K temperature range for pressure changes of 173(3), 260(8) and 390(12) MPa.

The isothermal entropy changes ($\Delta S_T$) associated to the barocaloric experiments can be indirectly determined using the following thermodynamic relation:[42]

$$\Delta S_T (T, \Delta p) = - \frac{c_p(T)}{T} \Delta T_S(T, \Delta p) \qquad (2)$$

where $c_P(T)$ is the specific heat at constant pressure. In our study, we obtained $c_P(T)$ from differential scanning calorimetry (DSC) measurements. $\Delta S_T$ obtained by this method for the 88 and 100 wt% samples is shown in Fig. 3, where $\Delta S_T$ is plotted as a function of temperature, within the 223-333 K range and for $\Delta p$ = 173, 260 and 390 MPa. Considering this calculation was based on the directly measured $\Delta T_S$ (Fig. 2a) and $c_p(T)$ presents a linear behavior within the considered temperature range, we can verify the qualitative behavior is the same in both data sets. For 390 MPa, the maximum values of ~95 J.kg$^{-1}$.K$^{-1}$ (WTR 88 wt%) and ~75 J.kg$^{-1}$.K$^{-1}$ (WTR 100 wt%) are comparable to the giant values obtained for VNR[31] and PDMS.[43]

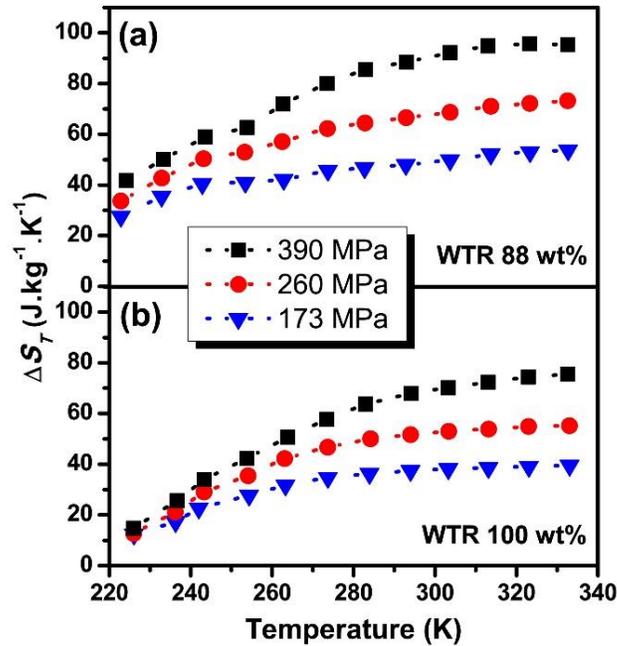

**Figure 3.** Isothermal entropy change as a function of temperature calculated using eq. 2 and $\Delta T_S$ data. The curves correspond to the WTR 88 wt% (a) and 100 wt% (b) samples, within the 223-333 K temperature range for pressure changes of 173(3), 260(8) and 390(12) MPa.

Performance parameters

Following the normalization of $i$-caloric effects recently proposed,[44] we calculated the normalized $\Delta T_S$ ($|\Delta T_S/\Delta p|$) and plotted against temperature (Fig. 4a) for VNR, WTR 88 and 100 wt% samples, in order to provide a proper evaluation of their barocaloric performance. It is worth mentioning that the best materials are located at the top right corner of the plot, the materials exhibiting high normalized values with high temperature change. One can see the dataset corresponding to VNR presents $|\Delta T_S/\Delta p|$ values consistently higher than those of WTR-based samples. Nevertheless, the values reached by the WTR 100 wt% (55.6 K.GPa$^{-1}$ at 333 K) are still remarkable, comparable or superior to other barocaloric materials in literature, as displayed in the graph.[32,45–48]

It is also important to analyze and compare the normalized refrigerant capacity (NRC) for barocaloric effect, which is defined as:[31,33]

$$NRC\ (\Delta T_{h-c}, \Delta p) = |\frac{1}{\Delta p} \int_{T_{cold}}^{T_{hot}} \Delta S_T(T, \Delta p) dT| \qquad (3)$$

where $\Delta T_{h-c}$ is the difference between the hot reservoir ($T_{hot}$) and the cold reservoir ($T_{cold}$). Fig. 4b shows the NRC curves calculated for $\Delta p$ = 173 MPa, using the $\Delta S_T$ determined from eq. (2). We have fixed the hot reservoir at 315 K for all samples. The initial values in the three curves are very close, but this difference increases for higher $\Delta T_{h-c}$. The maximum NRC value for WTR samples are lower than VNR at $\Delta T_{h-c}$ = 50 K, but once again higher than other materials with giant barocaloric effect.[32,45,48]

The whole analysis developed in this study reveals an encouraging scenario involving the use of waste rubbers in solid-state cooling applications. The values observed for both $\Delta T_S$ and $\Delta S_T$ and for the performance parameters, $|\Delta T_S/\Delta p|$ and NRC, clearly demonstrate WTR-based samples exhibit the required characteristics to act as refrigerants in barocaloric devices, surpassing most of the best barocaloric materials in literature.

An additional advantage in terms of performance obtained from WTR is its faster thermal exchange in comparison with VNR. A good way to evaluate the thermal exchange is to analyze the temperature-time data of our direct $\Delta T_S$ measurements and calculate, using Eq. 1 (Materials and Methods), the time constant ($\tau$) of the curve after the pressure is released.[41] Fig. 5 shows the time-constant ($\tau$) vs. WTR content for $\Delta p$ = 173 MPa at T = 273 K. We observe a significant decrease of the time constant ($\tau$) as the WTR content is increased in the samples. The time-constant drops from 25.6 s, for VNR, to 14.4 s, for WTR. But already at 88 wt% of WTR the value is 16.5 s. Taking other temperature, for the same pressure change (Table A2, Appendix), the average $\tau$ decrease in WTR 100 wt% sample, in comparison with pure VNR, is 11.1 s (~48 % lower). These results show that the heat flows faster from WTR during the barocaloric process. This effect can be explained by the relatively high concentration of carbon black[49,50] and other fillers in WTR powder. Bearing in mind the slow thermal exchange is still a challenge for elastomeric materials in view of cooling applications, the perspective of using WTR as refrigerant would concomitantly address this issue, besides its eco-friendly appeal. Finally, the particle size of WTR powder employed in our study was not controlled, and the VNR+WTR blends did not receive any post-synthesis processing. Nevertheless, several studies show the precise control of particle size and blending conditions can highly improve the characteristics of the synthesized blends.[4,17] Thus, we can expect even better caloric performances of VNR+WTR blends in further studies involving a systematic optimization of synthesis and processing.

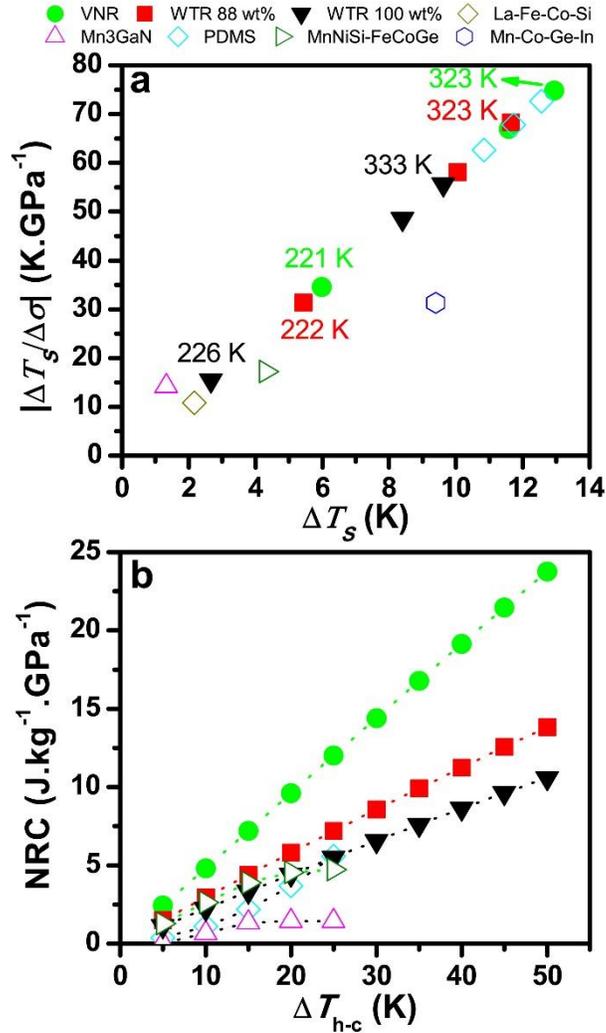

**Figure 4.** Performance parameters for VNR, WTR 88 and 100 wt% samples. (a) $|\Delta T_S/\Delta p|$ vs. $\Delta T_S$. The solid symbols correspond to the measured data for $|\Delta p| = 173$ MPa; the maximum and minimum temperatures of these data sets are indicated. The open symbols represent the following values from literature: PDMS ($|\Delta p| = 173$ MPa),[32] Mn$_3$GaN ($|\Delta p| = 93$ MPa),[45] La-Fe-Si-Co ($|\Delta p| = 200$ MPa),[46] Mn-Co-Ge-In ($|\Delta p| = 300$ MPa),[47] and MnNiSi-FeCoGe ($|\Delta p| = 250$ MPa).[48] (b) Normalized refrigerant capacity curves as a function of $\Delta T_{h-c}$, for $|\Delta p| = 173$ MPa (solid symbols). The hot reservoir was fixed at 315 K for the three measured samples. Values from literature (open symbols): PDMS (T$_{hot}$ = 315 K and $|\Delta p| = 173$ MPa), Mn$_3$GaN (T$_{hot}$ = 295 K and $|\Delta p| = 139$ MPa), and MnNiSi-FeCoGe (T$_{hot}$ = 335 K and $|\Delta p| = 270$ MPa). The dotted lines are guides for the eyes.

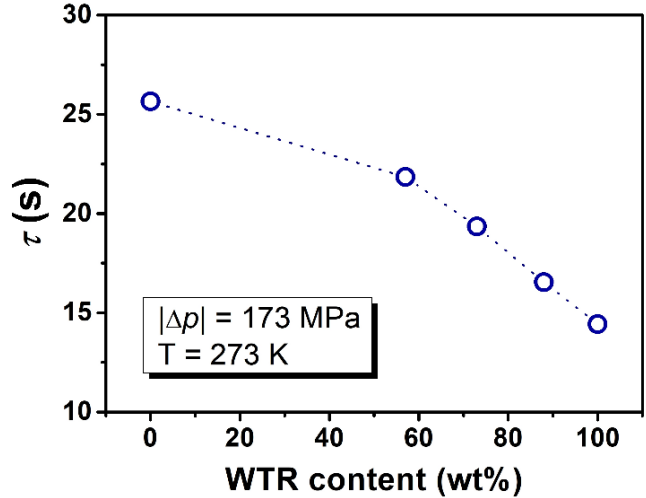

**Figure 5.** Time constant $\tau$ as function of the WTR content for $|\Delta p|$ = 173 MPa and T = 273 K. The values of $\tau$ were obtained from the direct $\Delta T_S$ measurements.

## CONCLUSIONS

A systematic investigation of the barocaloric characteristics of polymer blends made of VNR and WTR revealed that WTR-based samples present a slight reduction in the barocaloric effect in comparison with pure VNR. Nevertheless, the measured values are still within the giant barocaloric range: the maximum $\Delta T_S$ is 23.2 K (57 wt% WTR, at 333K for $\Delta p$ = 390 MPa) and the maximum $\Delta S_T$ is 95 J.kg$^{-1}$.K$^{-1}$ (88 wt% WTR, at 333 K for $\Delta p$ = 390 MPa). Also, the performance parameters are impressive, comparable or better than several barocaloric materials reported in the last years. In addition, WTR samples exhibit a faster thermal exchange than VNR, which may represent a great advantage for barocaloric cooling devices. All these findings evidence the promising potential of WTR in view of solid-state cooling applications, fostering a new alternative for recycling of waste tire rubber, besides contributing significantly to the field of sustainable energy technology.

ACKNOWLEDGMENTS

The authors acknowledge financial support from FAPESP (project number 2016/22934-3), CNPq and CAPES. The authors also thank LNLS and CNPEM.

APPENDIX: ADDITIONAL FIGURES AND TABLES

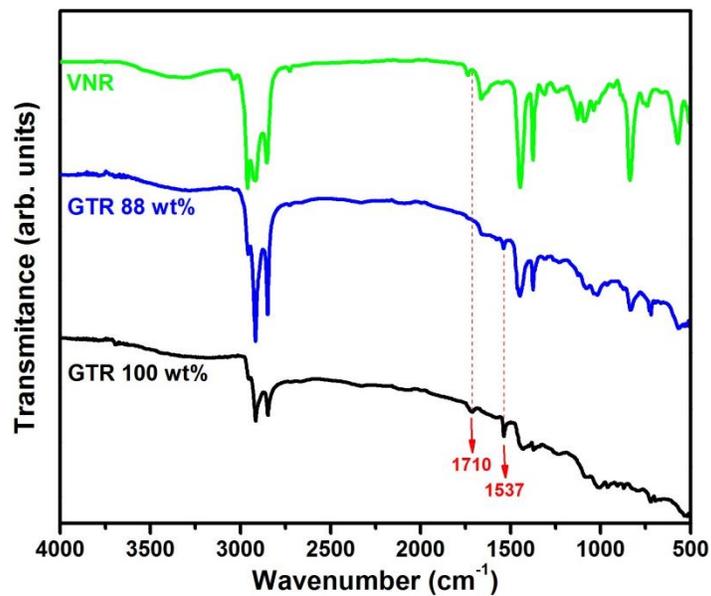

**Fig. A1**: FTIR spectra for pure WTR, WTR 88 wt% polymer blend and VNR.

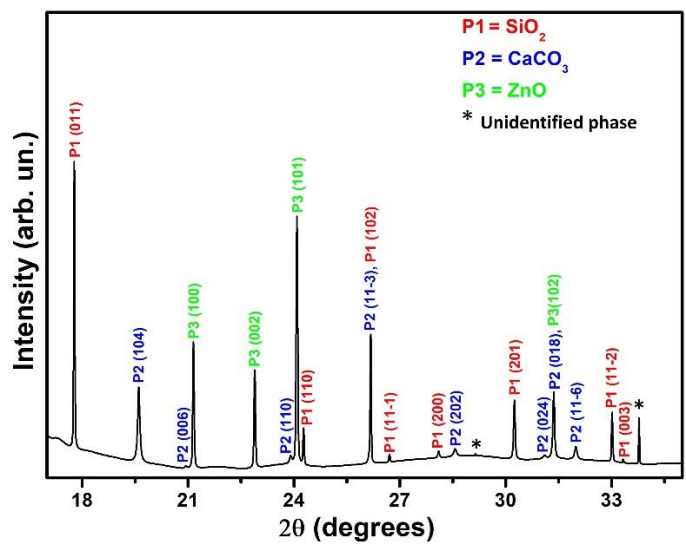

**Fig. A2**: X-ray diffraction pattern for pure WTR, with phase identification. X-ray wavelength: 1.033 Å.

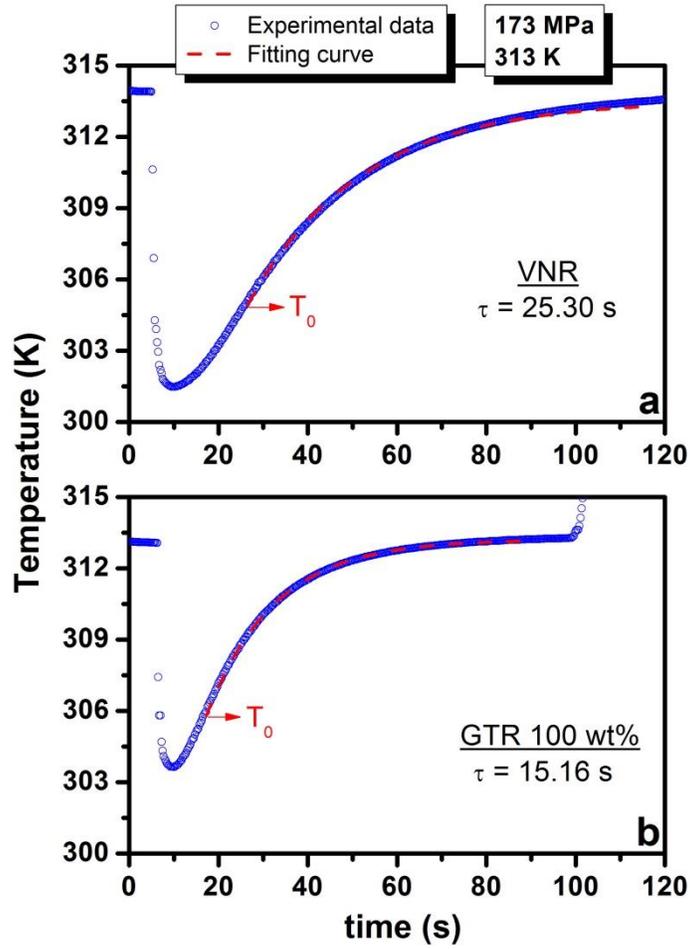

**Fig. A3**: Temperature *vs.* time experimental curves for (a) VNR and (b) pure WTR during a decompression cycle at 313 K for $\Delta p = 173$ MPa (blue open circles). The red dashed lines correspond to the fittings of data using eq. 1, where the initial temperature used in the model ($T_0$) and the calculated time constants ($\tau$) are indicated.

**Table A1**: Size distribution of granules in WTR powder, determined by optical microscopy.

| Diameter (μm) | Frequency (%) |
|---|---|
| 0 - 100 | 56.46 |
| 100 - 200 | 21.79 |
| 200 - 300 | 9.04 |
| 300 - 400 | 4.68 |
| 400 - 500 | 2.77 |
| > 500 | 5.26 |

**Table A2**: τ values calculated for VNR and pure WTR from temperature *vs.* time experimental curves. The relative gain corresponds to the percentage decrease in τ for WTR in comparison with VNR, at the same experimental conditions.

| Temperature (K) | $\tau_{WTR}$ (s) | $\tau_{VNR}$ (s) | Relative gain (%) |
|---|---|---|---|
| 333 | 17.94 | 28.47 | 36.99 |
| 323 | 19.28 | 24.68 | 21.86 |
| 313 | 15.16 | 25.30 | 40.07 |
| 303 | 13.89 | 22.24 | 37.57 |
| 293 | 12.15 | 24.18 | 49.78 |
| 283 | 11.92 | 24.88 | 52.12 |
| 273 | 14.42 | 25.56 | 43.58 |
| 263 | 10.36 | 20.13 | 48.55 |
| 253 | 6.47 | 20.42 | 68.31 |
| 243 | 8.35 | 19.99 | 58.21 |
| 233 | 7.37 | 23.38 | 68.48 |